# Strong Phonon-Cavity Coupling and Parametric Interaction in a Single Microcantilever under Ambient Conditions

*Qibin Zeng, Kaiyang Zeng**

Department of Mechanical Engineering,

National University of Singapore,

9 Engineering Drive 1, 117576, Singapore

*To whom correspondence should be addressed: mpezk@nus.edu.sg

**ORCID iDs**

Kaiyang Zeng 0000-0002-3348-0018

## Abstract

Parametrically tuning the oscillation dynamics of coupled micro/nano-mechanical resonators through a mechanical pump scheme has recently attracted great attentions from fundamental physics to various applications. However, the special design of the coupled resonators and low dissipation operation conditions significantly restrict the wide application of this tuning technique. In this study, we will show that, under ambient conditions, mechanical pump can parametrically control the oscillation dynamics in a *single* commercial microcantilever resonator. A strong phonon-cavity coupling with cooperativity up to ~398 and normal-mode splitting are observed in the microcantilever. The strong parametric interaction of the phonon-cavity coupling enables using mechanical pump to achieve a 43 dB (3 dB) parametric amplification (cooling). By utilizing mechanical pump, the force sensitivity and signal-to-noise ratio of the frequency-modulation Kelvin Probe Force Microscopy can be significantly improved in the ambient environment. Furthermore, both single-mode and two-mode thermomechanical noise squeezing states can be created in the microcantilever *via* applying mechanical pump.

# 1. Introduction

Detecting weak classical force and tiny mass underpins a variety of modern high-precision characterization technologies, such as gravity wave detection [1], Atomic Force Microscopy (AFM) [2], Kelvin Probe Force Microscopy (KPFM) [3], Magnetic Resonance Force Microscopy (MRFM) [4] and resonant mass sensing [5], and many others. For most of the force or mass measurements, the signals of interest are extracted from the motion of a mechanical resonator sensor [1, 6, 7], such as a microcantilever, where the ultimate detection sensitivity and signal-to-noise ratio (SNR) are commonly dominated by the oscillation dynamics [8, 9]. Driven by the continuous demands of better detecting sensitivity and SNR [8-11], numerous methods about tuning the oscillation dynamics of the mechanical resonator to optimize its sensitivity and SNR have been proposed, including feedback-based quality ($Q$) factor control [11-13], optomechanical pump [14-16], degenerate parametric pump [8, 17, 18] and mechanical pump [10, 19-22], *etc*. Among these methods, mechanical pump has received increasing attentions in recent years due to its unique characteristics [9, 23-25]. The mechanical pump is a non-degenerate parametric tuning method based on the energy transfer between two coupled mechanical modes which can be provided from one single resonator [19, 26, 27] or two coupled resonators [20, 23, 28]. By coupling the two mechanical modes, mechanical pump can imitate the optomechanical pump process to tune the oscillation dynamics of the mode, in which the electromagnetic mode (optical or microwave cavity) is replaced by a mechanical mode (phonon cavity) in the resonator [9, 19, 20]. Therefore, similar to the optomechanical pump technique [14], the mechanical pump also allows to amplify or cool the resonance (i.e., tune the effective $Q$ factor and temperature) and squeeze the thermomechanical noise of the resonator [9, 20, 26, 27, 29]. Although mechanical pump has been extensively demonstrated recently, nearly all of these studies are based on specially fabricated resonators which are operated in low dissipation environments, i.e., vacuum or even low temperature [9, 19-23, 25-34]. This is because effective mechanical pump requires a strong phonon-cavity coupling, the low dissipation environment and specific resonator structure can significantly promote the generation of such strong coupling [26, 29, 35]. Obviously, the requirements of low dissipation conditions and special fabrication of the resonator have fundamentally limited the broad application of the mechanical pump. Therefore, exploring the phonon-cavity coupling and mechanical pump of the mechanical resonator sensors, especially those commercially available resonators, under high dissipative environments will be of great importance for a wide range of high-precision measurement applications [5, 11, 36-40].

In this study, we report that, under ambient environment (*i.e.*, in air with room temperature), strong phonon-cavity coupling and efficient mechanical pump can be achieved in a *single*, commercially available microcantilever. By driving a commercial microcantilever-based AFM probe, we observe a strong phonon-cavity coupling with cooperativity up to ~398 between the $1^{st}$ flexural and $2^{nd}$ torsional modes of the microcantilever as well as the normal-mode splitting in both two modes, these observations can be well described by a parametrically coupled two-oscillator model. Due to the strong parametric interaction between the two modes, a 43 dB (3 dB) parametric amplification (cooling) is also achieved on the $1^{st}$ flexural mode *via* applying mechanical pump. More importantly, we unambiguously demonstrate that, by utilizing the mechanical pump, the force sensitivity and SNR of the frequency-modulation KPFM (FM-KPFM) can be significantly improved in the ambient environment. In addition, it is clearly shown that both single-mode and two-mode thermomechanical noise squeezing states can be created in a *single* microcantilever by applying the mechanical pump.

## 2. Results and Discussion

The microcantilever used in this study is the widely used Pt-Ir coated AFM probe (PPP-CONTSCPt, Nanosensors) with a nominal force constant and fundamental resonance frequency of ~0.2 N/m and ~25 kHz respectively. A probe holder with a piezo transducer is used to apply mechanical pump and drive the torsional mode of the microcantilever. To drive the flexural mode and demonstrate the mechanical pump in FM-KPFM, the microcantilever is kept near the surface of a conductive Pt substrate and an AC voltage is applied between the tip and substrate to induce an electrostatic excitation on the microcantilever. The vibration of the microcantilever is detected by the optical lever system of an AFM (SPA400, Seiko Instruments Inc., Japan) operated in ambient environment at room temperature. Figure 1a schematically shows the experimental set-up used for this study. Figure 1b and 1c show the measured frequency responses of the 1$^{st}$ flexural mode (mode 1) and 2$^{nd}$ torsional mode (mode 2) of the microcantilever respectively, which give the resonance frequencies ($\omega_1 = 2\pi \times 33.268$ kHz, $\omega_2 = 2\pi \times 932.630$ kHz), $Q$ factors ($Q_1 = 110.8$, $Q_2 = 532.6$) and dissipation rates ($\gamma_1 = \omega_1 / Q_1 = 2\pi \times 300$ Hz, $\gamma_2 = \omega_2 / Q_2 = 2\pi \times 1781$ Hz) of the two modes. Figure 1d shows the mode shapes of mode 1 and 2 obtained by finite element simulation (with approximate geometry sizes), and it is found that the experimentally measured eigenfrequencies (33.268 kHz and 932.630 kHz) can well match the simulation results (32.3 kHz and 979.3 kHz).

The strong phonon-cavity coupling in the AFM microcantilever is first observed from a surprising enhancement or even self-oscillation of the 1$^{st}$ flexural mode when the microcantilever is mechanically driven by the holder transducer around a frequency (~965 kHz) much higher than $\omega_1$, and similar phenomena can also be observed on multiple other PPP-CONTSCPt microcantilevers. Further experiments on these microcantilevers show that the resonance $Q$ factors of the 1$^{st}$ flexural mode can be significantly amplified by a factor of more than 100 *via* this scheme, thereby leading to a surprisingly high $Q$ factor of more than 10000 on such type of soft (~0.2 N/m) AFM probe in the ambient environment. In addition to the resonance amplification, it is also observed that the resonance of the 1$^{st}$ flexural mode can be suppressed (i.e., decrease the $Q$ factor) when the microcantilever is mechanically driven at another higher frequency (~899 kHz), implying that the resonance $Q$ factor of the microcantilever now can be bi-directionally tuned with this special mechanical excitation method. Further analyses show that, these intriguing phenomena and resonance tuning scheme are exactly above introduced strong phonon-cavity coupling and mechanical pump, which will be analyzed and discussed in detail in the following sections.

To confirm the phonon-cavity coupling and parametric interaction between the two modes in a single microcantilever, we firstly apply a weak electrostatic excitation at a driving frequency ($\omega_d$) which is near $\omega_1$ to the microcantilever, and at the same time, the microcantilever is mechanically pumped on the red-detuned sideband (red or anti-Stokes pump) at the pump frequency of $\omega_p$ which is near the difference frequency of the two modes, i.e., $\omega^- = \omega_2 - \omega_1 = 2\pi \times 899.362$ kHz (red sideband). During the measurement, $\omega_d$ and $\omega_p$ are swept within a narrow frequency region, and the frequency responses of mode 1 and mode 2 are simultaneously detected at the frequencies of $\omega_d$ and $\omega_{d+p} = \omega_d + \omega_p$, respectively. Figure 2a and 2b show the measured frequency responses of the mode 1 under two pump amplitudes of 0.1 $V_{pp}$ and 0.3 $V_{pp}$, respectively. Figure 2c and 2d show the simultaneously measured frequency responses of the mode 2 (at the frequency of $\omega_{d+p}$), in which are corresponding to the amplitude shown in Figure 2a and 2b, respectively. Obviously, when a small red pump is applied to the microcantilever, the amplitude decreases in mode 1 (Figure 2a) while increases in mode 2 (Figure 2c) when sweeping with the frequency of $\omega_p$, and a noticeable parametric deamplification can be observed in the mode 1 at the frequency of $\omega_p \approx \omega^-$. This phenomenon can be understood by the energy transfer diagram shown in Figure 1d. Under weak probe of mode 1 and red pump, phonons can be created in mode 2 at the expense of probe phonons in mode 1 and pump phonons *via* one-phonon adsorption process: $\hbar\omega_1 + \hbar\omega_p \rightarrow \hbar\omega_2$ (i.e., process "1" in Figure 1d) [23], thus the vibration of mode 1 is attenuated while a vibration in mode 2 is excited and enhanced with the increasing phonon reaction. With the red pump amplitude is further increased, evident normal-mode splitting with the avoided-crossing characteristic can be observed in both mode 1 (Figure 2b) and mode 2 (Figure 2d), in which clearly demonstrates the strong coupling between the two modes [25, 26]. The emergence of strong coupling indicates that the rate at which the phonons generated in mode 1 and 2 begins to exceed their rate of decay from both modes [29, 35]. Similarly, to probe the oscillation dynamics of mode 2, a weak mechanical excitation at $\omega_d$ near $\omega_2$ is applied to the microcantilever and at the same time, the same red pump is applied. The frequency response of the mode 2 monitored at $\omega_d$ and the corresponding response of mode 1 monitored simultaneously at the frequency of $\omega_{d-p} = \omega_d - \omega_p$ are shown in Figure 2i,j and 2k,l, respectively. When red pump is small, a mechanically induced transparency can be observed in mode 2 (Figure 2i), while at the position of transparency, the amplitude of mode 1 is increased instead (Figure 2k). The energy transfer process shown in Figure 1d can be used to interpret this phenomenon. With weak probe in mode 2 and red pump, phonons are converted from mode 2 to mode 1 and pump sideband *via* the reverse emission process: $\hbar\omega_2 \to \hbar\omega_1 + \hbar\omega_p$ (i.e., process "2" in Figure 1d) [20, 23], thereby causing the transparency in mode 2 and at the same time, vibration in mode 1. If further increase the amplitude of the red pump, an obvious normal-mode splitting can be observed in mode 2 (Figure 2j), further confirming the existence of strong coupling. Note that the frequency responses of the mode 1 and 2 are different under the weak red pump. The emergence of parametric deamplification in mode 1 and mechanically induced transparency in mode 2 indicates that in this phonon-cavity coupling system, mode 2 plays the role of phonon cavity [20, 27], thus the dynamical back-action of the phonon cavity can be employed to cool the thermomechanical oscillation of mode 1 [20, 27], and this will be discussed later.

The coupling and parametric interaction between mode 1 and 2 are further investigated under blue-detuned sideband pump (blue or Stokes pump) by applying a mechanical pump (7 $V_{pp}$) at a pump frequency of $\omega_p$, in which is near the sum frequency of the two modes, i.e., $\omega^+ = \omega_1 + \omega_2 = 2\pi \times 965.898$ kHz (blue sideband), to the microcantilever. Firstly, a weak electrostatic excitation at the driving frequency of $\omega_d$ near the $\omega_1$ is applied to probe mode 1. Figure 3a and 3b show the simultaneously measured frequency responses of the mode1 (detected at $\omega_d$) and mode 2 (detected at $\omega_{p-d} = \omega_p - \omega_d$), respectively. It is clear that both mode 1 and mode 2 show evident parametric amplification under the blue pump, and the vibration of mode 1 and 2 are enhanced synchronously. This result can be understood from the process "3" shown in Figure 1d, where the pump phonons convert to the phonons in both mode 1 and 2 via the down-conversion process, i.e., $\hbar\omega_p \rightarrow \hbar\omega_1 + \hbar\omega_2$, thereby compensating the energy dissipation and leading to mechanical parametric amplification in both modes simultaneously [9, 26, 31]. Similar results can be obtained when mode 2 is weakly probed by a mechanical drive at the frequency $\omega_d$ near the $\omega_2$, and the frequency responses of the mode 2 (detected at $\omega_d$) and mode 1 (detected at $\omega_{p-d}$) are shown in Figure 3e and 3f, respectively. Note that the amplification demonstrated here belongs to non-degenerate parametric amplification, which can be achieved in both modes simultaneously with evasion of the phase influence. This is obviously different with the widely used degenerate parametric amplification which is phase sensitive and requires phase information beforehand [8, 9, 25, 41].

To understand the underlying two-mode interaction mechanisms, the oscillation dynamics of this system is modeled by a parametrically coupled two-oscillator model [20, 23, 42]. By introducing a phenomenological coupling Hamiltonian [22], the total Hamiltonian of the system is:

$$H = \sum_{n=1}^{2}\left[\frac{P_n^2}{2m_n} + \frac{m_n\omega_n^2 X_n^2}{2} - F_n X_n \cos(\omega_d t)\right] - \Gamma X_1 X_2 \cos(\omega_p t) \quad (1)$$

where the canonical coordinates $X_n$ and $P_n$ denote the position and conjugate momentum of the mode with effective mass $m_n$ and frequency $\omega_n$, respectively; $F_n$ is the amplitude of the weak probing force; the last term describes the linearized parametric coupling between mode 1 and 2, where $\Gamma$ is the parametric inter-mode coupling coefficient that is proportional to the pump amplitude [20, 29]. Applying Hamiltonian canonical transform to Eq. 1 and considering an additional damping term, the motion of the two coupled modes are then given by

$$\frac{d^2X_1}{dt^2} + \gamma_1 \frac{dX_1}{dt} + \omega_1^2 X_1 = \frac{1}{m_1}\left[F_1 \cos(\omega_d t) + \Gamma X_2 \cos(\omega_p t)\right]$$
$$\frac{d^2X_2}{dt^2} + \gamma_2 \frac{dX_2}{dt} + \omega_2^2 X_2 = \frac{1}{m_2}\left[F_2 \cos(\omega_d t) + \Gamma X_1 \cos(\omega_p t)\right]$$
(2)

When mode 1 is weakly probed under red pump, we can decompose the $X_1(t)$ and $X_2(t)$ into rotating frame as $X_1(t) = \mathrm{Re}[A_1 \exp(i\omega_d t)]$ and $X_2(t) = \mathrm{Re}[A_2 \exp(i\omega_{d+p} t)]$, where $A_1$ and $A_2$ are slowly varying complex amplitudes of mode 1 and 2 respectively [23, 42]. Solving Eq. 2 in rotating frame and neglecting the time derivative items of $A_1$ and $A_2$ as well as the off-resonance items, the complex amplitudes of mode 1 and 2 can be obtained as following (details in Supplementary Material S1):

$$A_1 = \frac{2F_1 m_1(-\omega_{d+p}^2 + i\omega_{d+p}\gamma_2 + \omega_2^2)}{4m_1 m_2(\omega_d^2 - i\omega_d\gamma_1 - \omega_1^2)(\omega_{d+p}^2 - i\omega_{d+p}\gamma_2 - \omega_2^2) - \Gamma^2}$$
$$A_2 = \frac{-F_1 \Gamma}{4m_1 m_2(\omega_d^2 - i\omega_d\gamma_1 - \omega_1^2)(\omega_{d+p}^2 - i\omega_{d+p}\gamma_2 - \omega_2^2) - \Gamma^2}$$
(3)

With this method, the similar equations for all of the complex amplitudes can be obtained for the rest cases including probing mode 2 under red pump, probing mode 1 under blue pump and probing mode 2 under blue pump (Supplementary Material S1). Here, to qualitatively simulate the frequency responses of the two modes, we set the effective mass and probing force of both modes to be unity for the simplicity [20]. Then the frequency response of each mode is calculated from the real part of the complex amplitudes. According to Eq. 3, the frequency responses of the mode 1 and 2 can be calculated and the results are shown in Figure 2e,f and 2g,h respectively, where the coupling coefficient $\Gamma$ are set to the values of $6.2 \times 10^9$ (Figure 2e,g) and $1.86 \times 10^{10}$ (Figure 2f,h) in the calculation. Comparing Figure 2a-d and 2e-h, it is clear that the calculated frequency responses match the experimental results very well, confirming that the observed phenomenon is originated from the parametric coupling of the mode 1 and 2. For probing mode 2 under red pump, the measured frequency responses can be easily reproduced by Eq. S7 (Supplementary Material S1) and the calculated results are displayed in Figure 2m-p, with the values of $\Gamma = 6.2 \times 10^9$ (Figure 2m,o) and $1.86 \times 10^{10}$ (Figure 2n,p). Similarly, when the microcantilever is on blue pump, the frequency responses can be calculated by Eq. S9 (mode 1 is probed) and Eq. S11 (mode 2 is probed) (Supplementary Material S1), and the results are shown in Figure 3c,d and 3g,h respectively ($\Gamma = 7.6 \times 10^9$ for both). Obviously, the experimental and calculated frequency responses agree well with each other (Figure 3), indicating that the parametric amplification is unambiguously induced by the mode coupling.

On the other hand, the strength of the phonon-cavity coupling and parametric interaction can be easily tuned through the pump amplitude. Figure 4a shows the measured frequency response of the mode 1 (weakly probed near $\omega_1$) when pumped at $\omega_p = \omega^-$, and the simulated response (by Eq. S4) is shown in Figure 4b. With increasing the red pump amplitude, the resonance peak of mode 1 can be seen to decrease firstly, indicating the parametric deamplification. As the pump amplitude is further increased, the frequency response of the mode 1 splits into two well-resolved peaks with the separation given by $g/\pi$, where $g$ is the so-called coupling rate [20, 26]. Figure 4e shows the curves of coupling rate as a function of red pump amplitude, which are extracted from Figure 4a and 4b. It is clear that, when the resonance peak starts to split, the measured coupling rate shows the expected linear relationship with the pump amplitude, which agrees well with the calculation. At the highest pump amplitude used here (3.7 $V_{pp}$), the coupling rate $g = 2\pi \times 7220$ Hz, well satisfying the condition of strong coupling regime as $g > \gamma_2 \gg \gamma_1$ [20, 35]. This phonon-cavity coupling can be quantified by a figure of merit called cooperativity, defined as $C = 4g^2/(\gamma_1\gamma_2)$ (shown in Figure 4e), which is ~390 at the pump amplitude of 3.7 $V_{pp}$, indicating that ~390 cycles of energy transfer can be reached between the two modes before dissipating to the ambient environment [26]. The same measurement has been implemented on mode 2, and the results also show the expected trend of the linear splitting as well as the characteristics of a phonon cavity (Supplementary

Material S2) [20]. The parametric amplification of the two modes can also be directly controlled by the blue pump amplitude. Figure 4c shows the measured frequency response of mode 1 (weakly probed near $\omega_1$) when pumped at $\omega_p = \omega^+$, and the simulated response (by Eq. S9) is displayed in Figure 4d. Figure 4f shows the gains of resonance amplitude as a function of pump amplitude, which are extracted from Figure 4c and 4d. It can be seen that with increasing blue pump strength, the resonance linewidth decreases while the peak amplitude increases significantly, this is because the energy dissipation in mode 1 is growingly compensated by the pump energy thereby leading to increasing effective $Q$ factor. The same resonance amplification can be observed in mode 2 as well (Supplementary Material S2), which is highly in consistent with the prediction shown in the process "3" in Figure 1d.

Furthermore, to explore the limit of the mechanical pump, the thermal noise of the microcantilever is analyzed with the mechanical pump applied only. Figure 4g and 4h show the noise power spectral density (PSD) of the mode 1, $S_{11}(\omega)$, measured under blue ($\omega_p = \omega^+$) and red ($\omega_p = \omega^-$) pumps respectively. In Figure 4h, with increasing pump amplitude, the peak of $S_{11}(\omega)$ decreases firstly and then goes to splitting, which is similar with that shown in Figure 4a,b. The attenuation of $S_{11}(\omega)$ implies that mode 1 can be parametrically cooled by the dynamical back-action of the phonon cavity (*i.e.,* mode 2). Figure 4i shows the normalized temperature $T/T_0 = \int S_{11}(\omega)d\omega / \int S_{11}(\omega)\big|_0 d\omega$ at various pump amplitudes. It can be seen that the cooling efficiency increases with the pump amplitude, which can be understood from the phonon transfer between mode 1 and the phonon cavity [35]. The maximum cooling ratio attained here is ~2.0 (3 dB), implying that the effective temperature is decreased from room temperature (~298 K) to ~149 K by parametric cooling. At the maximum pump amplitude before entering parametric instability, the measured coupling rate reaches up to ~$2\pi \times 7290$ Hz, corresponding to a cooperativity ($C$) of ~398 which is much higher than that of many counterparts operated in vacuum, such as the graphene drum resonator ($C = 60$) [26] and GaAs-based mechanical resonators ($C = 5 \sim 18$) [20, 21, 23, 25, 31]. When mode 1 is on blue pump, $S_{11}(\omega)$ shows an increasing enhancement of resonance with improving pump strength (Figure 4g). The effective $Q$ of mode 1 ($Q_{\text{eff-1}}$) fitted from $S_{11}(\omega)$ *via* Eq. S18 (Supplementary Material S3) is displayed in Figure 4j. At the highest pump amplitude before reaching the region of instability, the $Q_{\text{eff-1}}$ increases from ~110 to as high as ~18200 while the amplitude is amplified by a factor of ~150 (43 dB), indicating that the $Q$ factor of such type of soft AFM microcantilever (~0.2 N/m) can be easily enhanced to more than 10000 in the ambient environment. When pump amplitude is further increased, the system ran into parametric instability leading to self-oscillation (mechanical lasing) of mode 1. Note that the mechanical pump demonstrated here does not introduce potential thermal effect (such as Joule heating [20, 29, 43] in electrically or photothermal effect [44] in optically pumped system) to the microcantilever, implying that the temperature thereby the resonance frequency of the microcantilever can keep constant with increasing pump strength (Supplementary Material S4).

So far, we have unambiguously demonstrated the tuning of oscillation dynamics by mechanical pump on a single microcantilever under ambient conditions. Next, we will show that this study can be immediately applied to KPFM to improve the force sensitivity and SNR. KPFM is an important yet powerful AFM-based surface potential characterization technique which has been extensively used to study the electrical properties of various materials at nanoscale [3, 45]. KPFM mainly has two operation mode, the frequency-modulation (FM) and amplitude-modulation (AM) mode [46]. FM-KPFM measures the electrostatic force gradient-induced resonance frequency shift to give the surface potential information, and it has been confirmed to have much higher accuracy and spatial resolution than that of the AM-KPFM [45, 46]. However, KPFM is conventionally preferred to operate in AM instead of FM mode in ambient or liquid environment, because the viscous damping induced small $Q$ significantly degrades the force sensitivity of the FM-KPFM measurements [47]. Although the feedback-based "$Q$ control" technique have been used to adjust $Q_{\text{eff}}$ in dynamic AFM [48, 49], this method is invalid in FM-KPFM. To this extend, the mechanical pump demonstrated here provides an ideal solution for the application of FM-KPFM in the high dissipation environment. Figure 5a shows the open-loop FM-KPFM frequency shift ($\Delta f$) measured with an alternating total potential difference ($\Delta V$) of 0 and 2 V under blue pump $\omega_p = \omega^+$ (experimental details in Supplementary Material S5). Obviously, with increasing blue pump amplitude, $\Delta f$ shows an increasingly apparent difference between the two states with $\Delta V = 0$ and 2 V, clearly implying an improved force thus potential sensitivity. To quantitively verify the improvement of the sensitivity, the $\Delta f$ difference ($\delta f$) at two potential states ($\Delta V = 0$ and $V_h$) is measured with various values of $V_h$, and the relative error, $\eta = \delta f / \Delta f$, is calculated and plotted as a function of $V_h$ (Figure 5b). It can be seen that, without blue pump, $\eta$ dramatically increases with decreasing $V_h$, indicating an inferior potential sensitivity. In contrast, at the same $V_h$, $\eta$ can be significantly minimized when blue pump is on. If setting $\eta = 100\%$ as the upper limit of the error, the minimum detectable potential here is ~0.06 V with the mechanical pump and this has shown significantly improvement comparing with the value of ~1.0 V for the FM-KPFM without applying mechanical pump. The improvement of the sensitivity is ~16 times by using the mechanical pump. Meanwhile, the SNR of the $\Delta f$ signal can be calculated by

$$SNR = 10\log_{10}\frac{\langle \Delta f^2 \rangle}{\langle \Delta f_{noise}^2 \rangle} \tag{4}$$

When $\Delta V = 2$ V, the obtained SNR are 7.8 dB and 39.0 dB for pump amplitude of 0 (pump off) and 12.7 $V_{pp}$ respectively (calculated from Figure S4), indicating that the SNR can be significantly enhanced by mechanical pump as well. The successful application of mechanical pump in FM-KPFM will unambiguously imply that this tuning method can be promoted to many other AFM-based techniques, such as the Non-Contact AFM [50], MRFM [4] and Electrostatic Force Microscopy [51], and at the same time, used for microcantilever-based resonant mass sensing [5, 39, 40].

Typically, for a resonator sensor with strong Brownian motion, such as the case studied here, the force sensitivity is ultimately limited by the thermomechanical noise regardless the resonance amplification technique is used [8]. Fortunately, the mechanical pump offers great capabilities to produce parametric cooling and thermomechanical squeezing state in the resonator, which can break the thermal noise limitation thus further improve the force sensitivity [8, 27, 29, 31]. The thermal noise displacement from both modes of the microcantilever can be decomposed into slowly varying inphase $x_n$ and quadrature $y_n$ components within a narrow bandwidth *via* $X_n = x_n \cos(\omega_n t) + y_n \sin(\omega_n t) (n = 1, 2)$ [8, 29]. When parametric cooling is implemented, the fluctuations of $x_n$ and $y_n$ will be uniformly cooled down leading to attenuation of the thermal noise in both two quadratures [52], this can be observed from Figure 5c where mode 1 is parametrically cooled by applying red pump ($\omega_p = \omega^-$, 0.2 $V_{pp}$). In comparison to uniform cooling, the thermomechanical squeezing allows noise in one quadrature to be reduced with a possibility of infinite reduction at the expense of increasing noise in the orthogonal quadrature [8, 52]. Figure 5d shows the steady-state noise squeezing of mode 1, which is created *via* a reservoir-engineering scheme [52, 53] by simultaneously applying red and blue pump at $\omega_p = \omega^-$ and $\omega^+$ respectively. Squeezing *via* reservoir-engineering has been extensively studied in optomechanical pump system [52-54], whereas it has seldom been demonstrated in pure mechanical pump system. The results shown in Figure 5d clearly indicate that the single-mode noise squeezing can be achieved *via* applying mechanical reservoir-engineering pump (i.e., applying the red and blues pumps at the same time; here the frequency and amplitude are: $\omega_p = \omega^-$, 0.25 $V_{pp}$ for red pump and $\omega_p = \omega^+$, 19.0 $V_{pp}$ for blue pump). Similar with that of the optomechanical pump [52, 53], the squeezing ratio can also be adjusted in the mechanical pump system by tuning the amplitude ratio of the red and blue pumps. Besides the single-mode squeezing, the mechanical pump also allows to form the noise squeezing across mode 1 and 2, *i.e.*, the two-mode squeezing, through the parametric down-conversion process (Figure 1d, process "3") [29, 34]. The self- ($\{x_1, y_1\}$ and $\{x_2, y_2\}$) and cross-quadratures distributions ($\{x_2, y_1\}$ and $\{x_1, y_2\}$) of the normalized noise data measured under the blue pump ($\omega_p = \omega^+$, 12.5 $V_{pp}$) are shown in Figure 5e and 5f respectively. It is obvious that Figure 5e,f reveal a two-mode thermal noise squeezing, where the noises of mode 1 and 2 exhibit symmetric solid-circle distributions in the self-quadratures while clear squeezed distributions can be observed in the cross-quadratures. The two-mode squeezing ratio can be tuned by simply changing the blue pump amplitude.

## 3. Conclusion

In this study, we have unambiguously demonstrated strong phonon-cavity coupling and parametric interaction in a *single* commercial microcantilever under ambient conditions. By driving the microcantilever in ambient atmosphere at room temperature, we observed a strong phonon-cavity coupling with cooperativity up to ~398 between mode 1 and 2 as well as the normal-mode splitting in both two modes, which can be well described by a parametrically coupled two-oscillator model. Due to the strong parametric interaction between these two modes, a 43 dB (3 dB) parametric amplification (cooling) is achieved on mode 1 *via* mechanical pump. Furthermore, we unambiguously show that the mechanical pump can be used to significantly improve the force sensitivity and SNR of the FM-KPFM in ambient environment. In addition, by applying mechanical pump, both single-mode and two-mode thermomechanical noise squeezing states can be created in the microcantilever. According to the results of this study, it can be expected that by exploiting the phonon-cavity coupling in mechanical resonators, the force sensitivity, SNR and thermomechanical noise level in those microcantilever-based characterization techniques, such as AFM-based scanning probe microscopies and resonant mass sensing, can achieve a substantial optimization *via* the mechanical pump scheme in the high dissipation environments.

# Figures

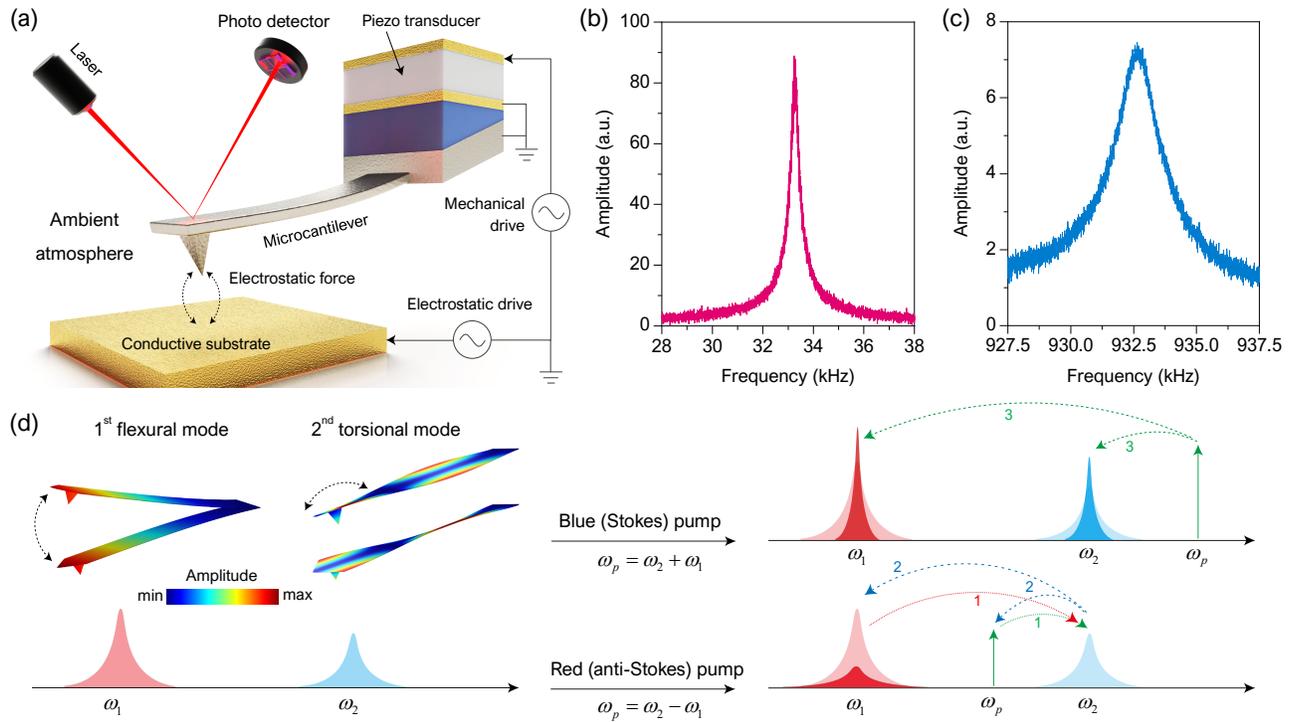

**Figure 1.** The single microcantilever system under test. (a) The experimental set-up for measuring the dynamic motion of the microcantilever. (b) The resonance curve of the 1st flexural mode and (c) the 2nd torsional mode. (d) Schematic illustration of the parametric pump in frequency space. Upper left shows the mode shape of the two coupled resonant mode; the curved arrows at the right side indicate the direction of energy transfer when the system is pumped at frequency $\omega_p$; and the labels "1", "2" and "3" denote that the energy transfers from mode 1 and pump to mode 2, from mode 2 to pump and mode 1, and from pump to mode 1 and 2, respectively.

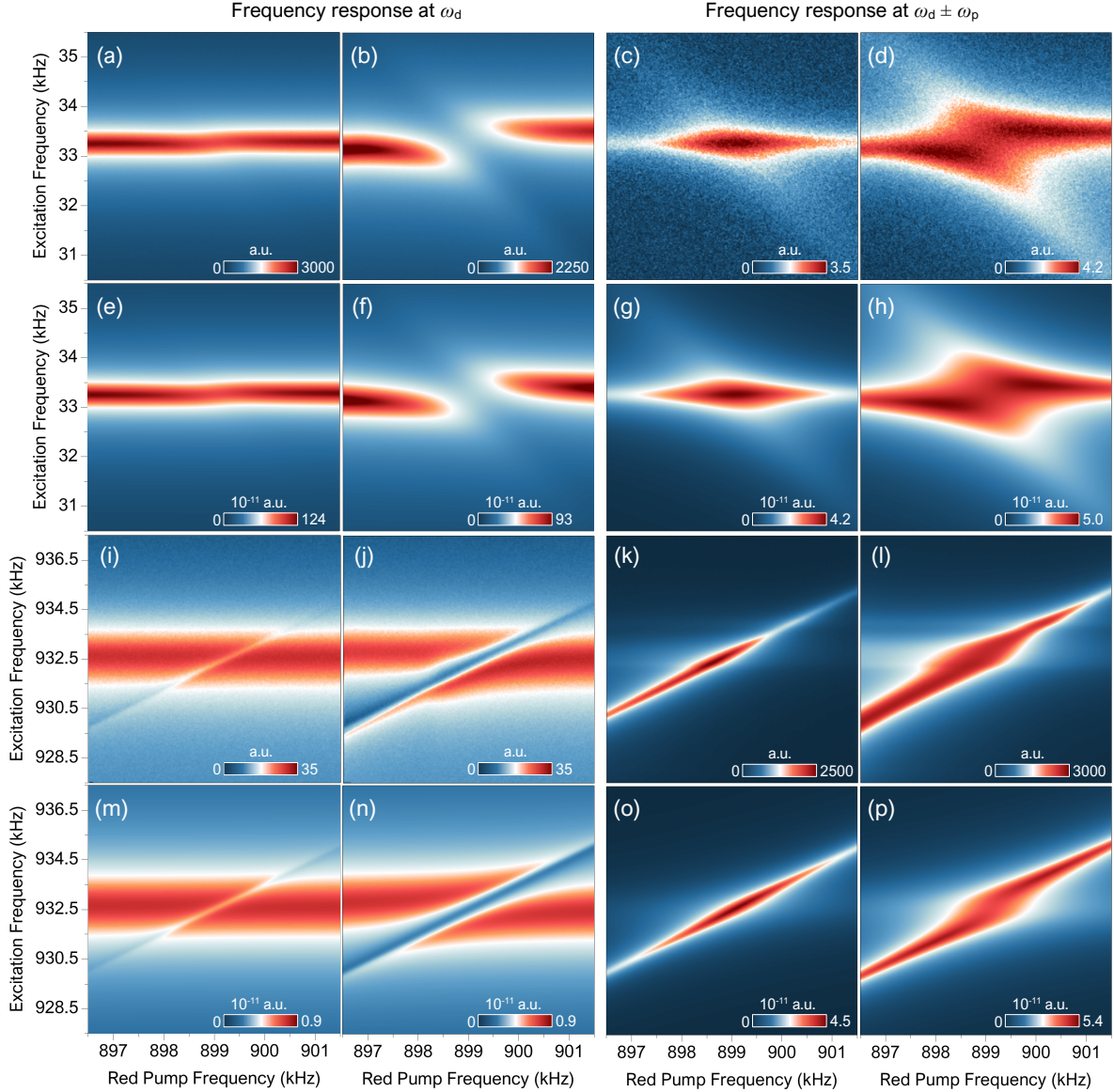

**Figure 2.** Parametric interaction and normal-mode splitting under red pump. (a, b) Frequency responses of mode 1 under red pump near the frequency $\omega^-$ with two pump amplitudes of 0.1 and 0.3 $V_{pp}$, respectively. (c, d) Simultaneously measured frequency responses of mode 2 corresponding to the pump frequency and amplitude of (a) and (b) respectively. (e-h) Calculated results corresponding to (a-d) respectively. (i, j) Frequency responses of mode 2 under red pump near the frequency $\omega^-$ with two pump amplitudes of 0.1 and 0.3 $V_{pp}$, respectively. (k, l) Simultaneously measured frequency responses of mode 1 corresponding to the pump frequency and amplitude of (i) and (j) respectively. (m-p) Calculated results corresponding to (i-l) respectively.

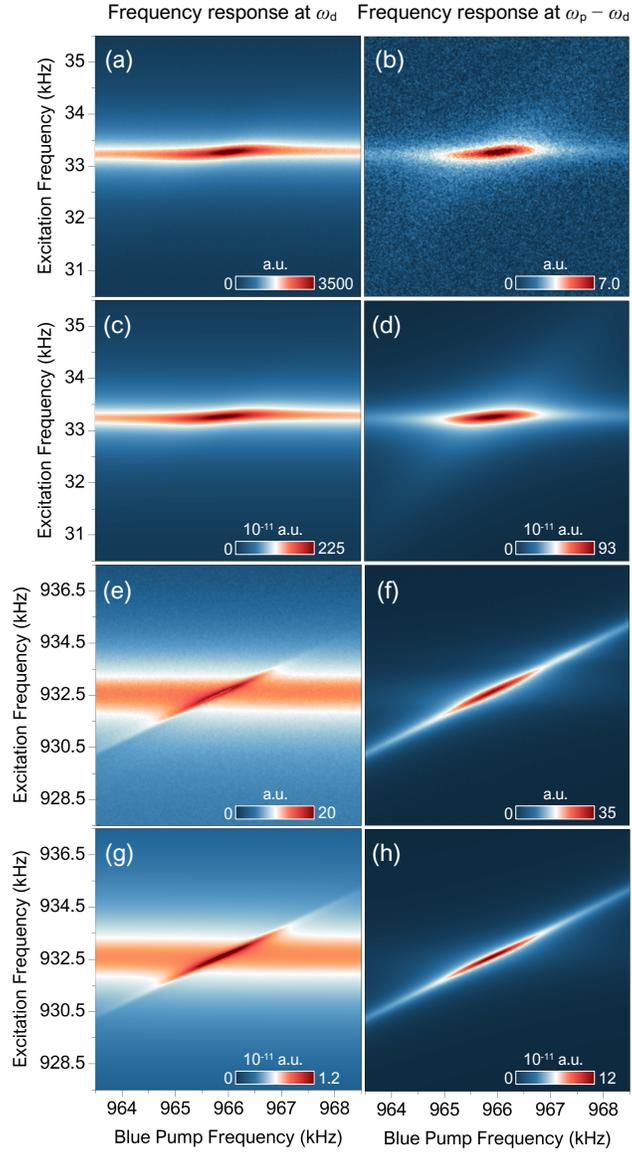

**Figure 3.** Parametric amplification under blue pump. (a) Simultaneously measured frequency responses of mode 1; and (b) mode 2 under blue pump near the frequency $\omega^+$ with pump amplitude of 7 $V_{pp}$. (c, d) Calculated results corresponding to the pump frequency and amplitude of (a) and (b) respectively. (e) Simultaneously measured frequency responses of mode 2; and (f) mode 1 under blue pump near the frequency $\omega^+$ with pump amplitude of 7 $V_{pp}$. (g, h) Calculated results corresponding to the pump frequency and amplitude of (e) and (f) respectively.

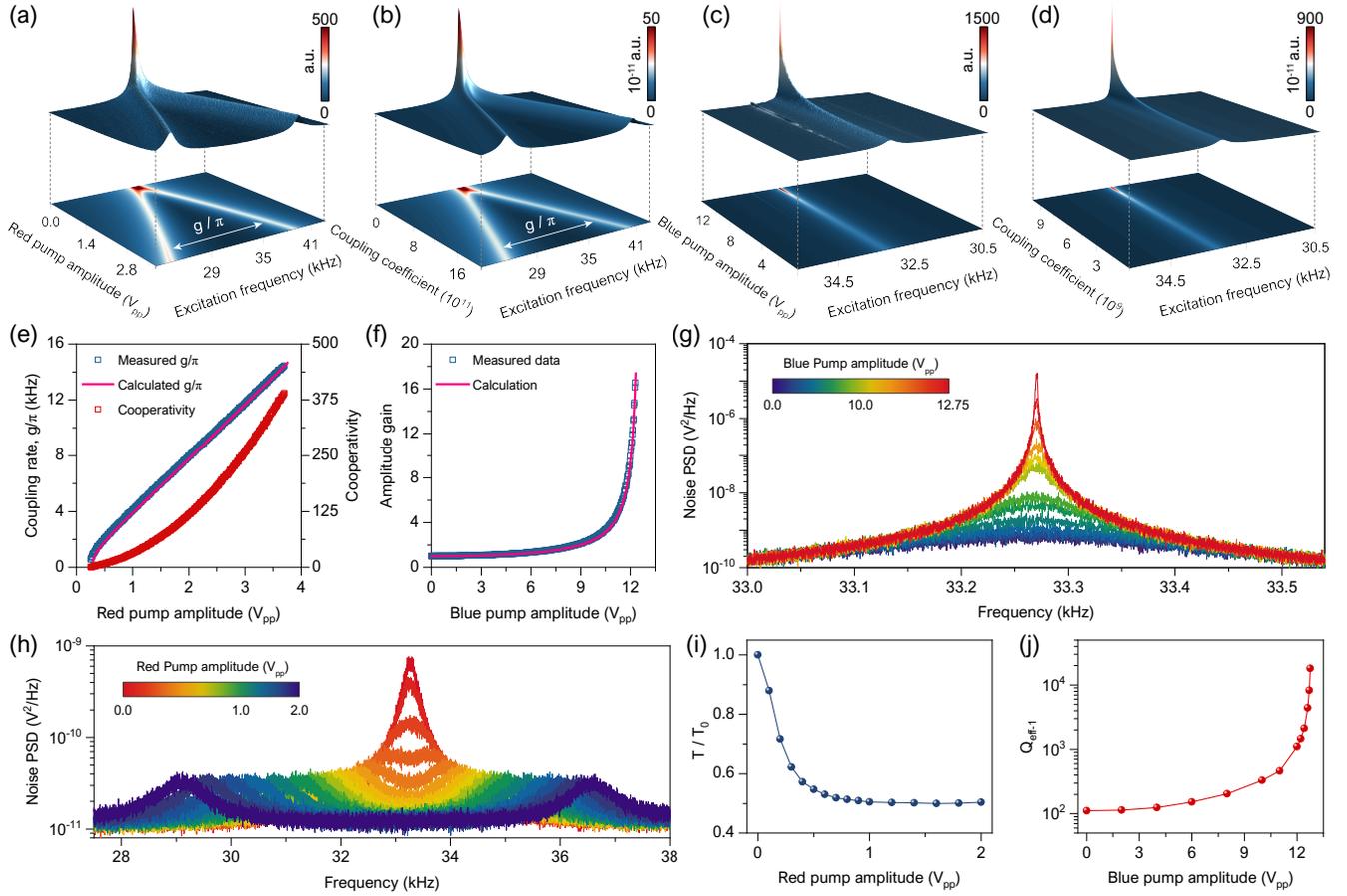

**Figure 4.** Tuning the strength of phonon-cavity coupling and parametric interaction. (a) Experimentally measured and (b) calculated frequency responses of mode 1 as a function of red pump amplitude and frequency. (c) Experimentally measured and (d) calculated frequency responses of mode 1 as a function of blue pump amplitude and frequency. (e) The coupling rate and cooperativity as a function of red pump amplitude. (f) The amplitude gain of mode 1 as a function of blue pump amplitude. (g, h) Noise PSD of mode 1 measured with different blue and red pump amplitudes respectively. (j) Normalized temperature of mode 1 as a function of red pump amplitude. (j) The effective $Q$ factor of mode 1 as a function of blue pump amplitude.

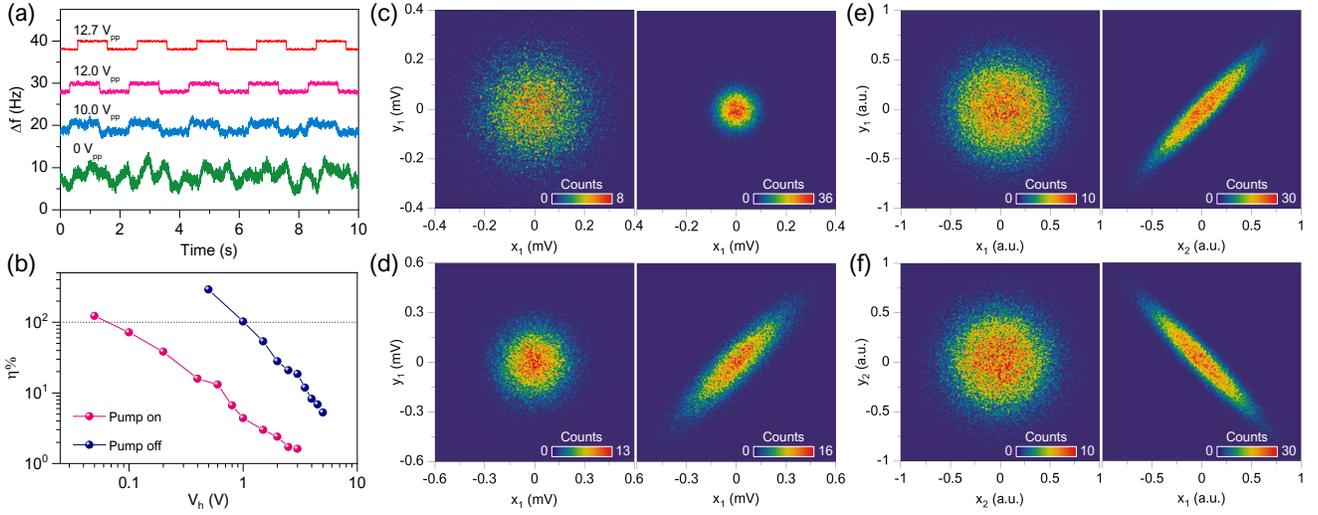

**Figure 5.** Application of mechanical pump under ambient conditions. (a) FM-KPFM frequency shift signal measured with different blue pump amplitudes, where all the curves are offset for clarity. (b) The relative error of $\Delta f$ as a function of applied $V_h$. (c) Phase-space histograms of the noise data $\{x_1, y_1\}$ in the absence (left) and presence (right) of parametric cooling. (d) Phase-space histograms of the noise data $\{x_1, y_1\}$ with (right) and without (left) applying reservoir-engineering pump. (e) Phase-space histograms of the normalized noise data $\{x_1, y_1\}$ (left) and $\{x_2, y_1\}$ (right) under blue pump. (f) Phase-space histograms of the normalized noise data $\{x_2, y_2\}$ (left) and $\{x_1, y_2\}$ (right) under blue pump.

# Acknowledgment

This work is supported by the Ministry of Education (MoE) Singapore through National University of Singapore (NUS) under the Academic Research Fund (AcRF) with funding number R-265-000-596-112. Q. Z. would like to acknowledge the scholarship support from MoE Singapore through NUS under AcRF of R-265-100-596-112 and Department of Mechanical Engineering, NUS. We would also like to thank Prof. Li Lu (NUS) for great support of the SPA400 AFM in this study.

**Supplementary Material:** Analytical solution of the parametrically coupled two-oscillator model, frequency response of mode 2 as a function of pump amplitude, effective Q factor under mechanical pump, resonance frequency of mode 1 as a function of blue pump amplitude, the measurement of FM-KPFM frequency shift under mechanical pump, FM-KPFM frequency shift signal for calculating SNR.

*Supplementary Material*

# Strong Phonon-Cavity Coupling and Parametric Interaction in a Single Microcantilever under Ambient Conditions

*Qibin Zeng, Kaiyang Zeng**

Department of Mechanical Engineering,

National University of Singapore,

9 Engineering Drive 1, 117576, Singapore

*To whom correspondence should be addressed: mpezk@nus.edu.sg

## S1. Analytical Solution of the Parametrically Coupled Two-Oscillator Model

### S1.1 Weakly probing mode 1 under red pump

Under the mechanical red pump with a frequency of $\omega_p$, when mode 1 is weakly probed by an electrostatic excitation $F_1\cos(\omega_d t)$, the motion of the two coupled modes can be described by

$$\frac{d^2X_1}{dt^2} + \gamma_1\frac{dX_1}{dt} + \omega_1^2 X_1 = \frac{1}{m_1}\left[F_1\cos(\omega_d t) + \Gamma X_2\cos(\omega_p t)\right]$$
$$\frac{d^2X_2}{dt^2} + \gamma_2\frac{dX_2}{dt} + \omega_2^2 X_2 = \frac{1}{m_2}\Gamma X_1\cos(\omega_p t) \quad (S1)$$

In this case, the motion of mode 2 is located at the frequency of $\omega_{d+p} = \omega_d + \omega_p$, and with introducing the complex amplitudes, the canonical position coordinates of mode 1 and 2 can be expressed as:

$$X_1(t) = \text{Re}\left[A_1(t)e^{i\omega_d t}\right]$$
$$X_2(t) = \text{Re}\left[A_2(t)e^{i\omega_{d+p} t}\right] \quad (S2)$$

Substituting Eq. S2 into Eq. S1, and ignoring the 2$^{nd}$-order time differential of complex amplitudes, the equations of motion can be expressed by the following equations:

$$(2i\omega_d + \gamma_1)\frac{dA_1(t)}{dt} + (i\omega_d\gamma_1 + \omega_1^2 - \omega_d^2)A_1(t) = \frac{1}{2m_1}\left[F_1 + \Gamma A_2(t)\right] + \frac{1}{2m_1}\left[F_1 e^{-2i\omega_d t} + \Gamma A_2(t)e^{2i\omega_p t}\right]$$
$$(2i\omega_{d+p} + \gamma_2)\frac{dA_2(t)}{dt} + (i\omega_{d+p}\gamma_2 + \omega_2^2 - \omega_{d+p}^2)A_2(t) = \frac{1}{2m_2}\Gamma A_1(t) + \frac{1}{2m_2}\Gamma A_1(t)e^{2i\omega_d t} \quad (S3)$$

At the steady state, $dA_1(t)/dt = dA_2(t)/dt = 0$, then by neglecting all the off-resonance components [1], the time-independent solution of Eq. S3 is given by:

$$A_1 = \frac{2F_1 m_1(-\omega_{d+p}^2 + i\omega_{d+p}\gamma_2 + \omega_2^2)}{4m_1 m_2(\omega_d^2 - i\omega_d\gamma_1 - \omega_1^2)(\omega_{d+p}^2 - i\omega_{d+p}\gamma_2 - \omega_2^2) - \Gamma^2}$$
$$A_2 = \frac{-F_1\Gamma}{4m_1 m_2(\omega_d^2 - i\omega_d\gamma_1 - \omega_1^2)(\omega_{d+p}^2 - i\omega_{d+p}\gamma_2 - \omega_2^2) - \Gamma^2} \quad (S4)$$

### S1.2 Weakly probing mode 2 under red pump

When mode 2 is weakly probed by a mechanical excitation $F_2\cos(\omega_d t)$, the motion of the two coupled modes now is described by:

$$\frac{d^2X_1}{dt^2} + \gamma_1\frac{dX_1}{dt} + \omega_1^2 X_1 = \frac{1}{m_1}\Gamma X_2\cos(\omega_p t)$$
$$\frac{d^2X_2}{dt^2} + \gamma_2\frac{dX_2}{dt} + \omega_2^2 X_2 = \frac{1}{m_2}\left[F_2\cos(\omega_d t) + \Gamma X_1\cos(\omega_p t)\right] \quad (S5)$$

In this case, the motion of mode 1 is located at the frequency of $\omega_{d\text{-}p} = \omega_d - \omega_p$, and with introducing complex amplitudes, the canonical position coordinates of mode 1 and 2 can be expressed by:

$$X_1(t) = \text{Re}\left[A_1(t)e^{i\omega_{d-p}t}\right]$$
$$X_2(t) = \text{Re}\left[A_2(t)e^{i\omega_d t}\right] \quad (S6)$$

Substituting Eq. S6 into Eq. S5 and using the similar process shown in case (1), the steady-state solution of the complex amplitudes is given by:

$$A_1 = \frac{-F_2\Gamma}{4m_1m_2(\omega_d^2 - i\omega_d\gamma_2 - \omega_2^2)(\omega_{d-p}^2 - i\omega_{d-p}\gamma_1 - \omega_1^2) - \Gamma^2}$$
$$A_2 = \frac{2F_2m_1(-\omega_{d-p}^2 + i\omega_{d-p}\gamma_1 + \omega_1^2)}{4m_1m_2(\omega_d^2 - i\omega_d\gamma_2 - \omega_2^2)(\omega_{d-p}^2 - i\omega_{d-p}\gamma_1 - \omega_1^2) - \Gamma^2} \quad (S7)$$

**S1.3 Weakly probing mode 1 under blue pump**

Under the mechanical blue pump with a frequency of $\omega_p$, when mode 1 is weakly probed by an electrostatic excitation $F_1\cos(\omega_d t)$, the motion of the two coupled modes can be described by Eq. S1. In this case, the motion of mode 2 is located at the frequency of $\omega_{p\text{-}d} = \omega_p - \omega_d$, and the canonical position coordinates of mode 1 and 2 can be expressed by:

$$X_1(t) = \text{Re}\left[A_1(t)e^{i\omega_d t}\right]$$
$$X_2(t) = \text{Re}\left[A_2(t)e^{i\omega_{p-d}t}\right] \quad (S8)$$

Substituting Eq. S8 into Eq. S1, the final steady-state solution of the complex amplitudes is given by:

$$A_1 = \frac{2F_1m_1(-\omega_{p-d}^2 + i\omega_{p-d}\gamma_2 + \omega_2^2)}{4m_1m_2(\omega_d^2 + i\omega_d\gamma_1 - \omega_1^2)(\omega_{p-d}^2 - i\omega_{p-d}\gamma_2 - \omega_2^2) - \Gamma^2}$$
$$A_2 = \frac{-F_1\Gamma}{4m_1m_2(\omega_d^2 + i\omega_d\gamma_1 - \omega_1^2)(\omega_{p-d}^2 - i\omega_{p-d}\gamma_2 - \omega_2^2) - \Gamma^2} \quad (S9)$$

**S1.4 Weakly probing mode 2 under blue pump**

Under the mechanical blue pump with a frequency of $\omega_p$, when mode 2 is weakly probed by a mechanical excitation $F_2\cos(\omega_d t)$, the motion of the two coupled modes can be described by Eq. S5. In this case, the motion of mode 1 is located at the frequency of $\omega_{p\text{-}d} = \omega_p - \omega_d$, and the canonical position coordinates of mode 1 and 2 can be expressed by:

$$X_1(t) = \text{Re}\left[A_1(t)e^{i\omega_{p-d}t}\right]$$
$$X_2(t) = \text{Re}\left[A_2(t)e^{i\omega_d t}\right] \quad (S10)$$

Substituting Eq. S10 into Eq. S5, the final steady-state solution of the complex amplitudes is given by:

$$A_1 = \frac{-F_2\Gamma}{4m_1m_2(\omega_d^2 + i\omega_d\gamma_2 - \omega_2^2)(\omega_{p-d}^2 - i\omega_{p-d}\gamma_1 - \omega_1^2) - \Gamma^2}$$

$$A_2 = \frac{2F_2 m_1(-\omega_{p-d}^2 + i\omega_{p-d}\gamma_1 + \omega_1^2)}{4m_1m_2(\omega_d^2 + i\omega_d\gamma_2 - \omega_2^2)(\omega_{p-d}^2 - i\omega_{p-d}\gamma_1 - \omega_1^2) - \Gamma^2}$$

(S11)

## S2. Frequency Response of Mode 2 as A Function of Pump Amplitude

The frequency responses of mode 2 as a function of the amplitude of the red and blue pumps are shown in Figure S1a and S1b respectively. Figure S1a shows the measured frequency response of the mode 2 (weakly probed near $\omega_2$) when pumped at $\omega_p = \omega^-$, and the calculated response is shown in Figure S1c. Figure S1b shows the measured frequency response of the mode 2 (weakly probed near $\omega_2$) when pumped at $\omega_p = \omega^+$, and the calculation result is displayed in Figure S1d.

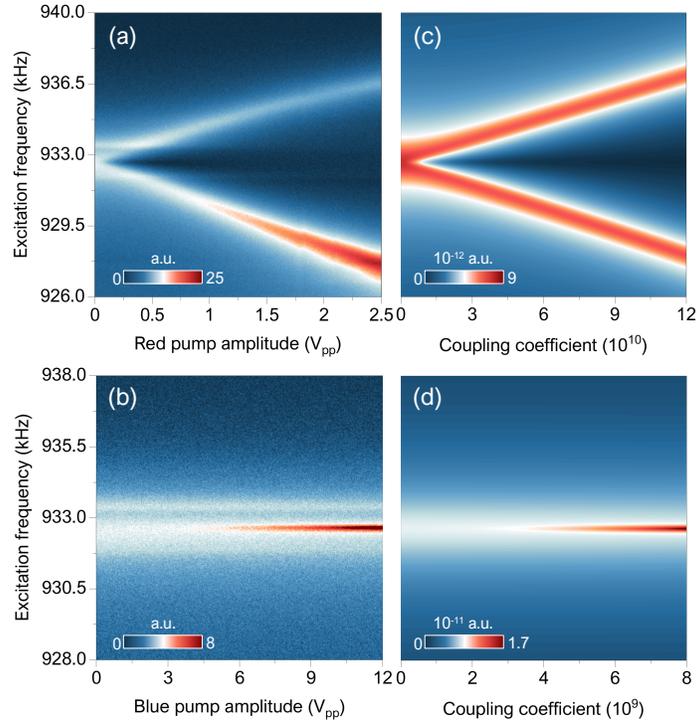

**Figure S1.** Frequency response of mode 2 as a function of pump amplitude. (a) Frequency response of mode 2 as a function of red and (b) blue pump amplitude. (c, d) Calculation results for (a, b) respectively.

## S3. Effective *Q* Factor under Mechanical Pump

Considering the only stimulation of thermomechanical noise force, $F_{th}$, the intrinsic equation of motion for a harmonic oscillator is given by:

$$m\ddot{z}(t) + \zeta \dot{z}(t) + kz(t) = F_{th} \tag{S12}$$

where $z(t)$, $m$, $\zeta$ and $k$ are the vibration displacement, lumped mass, linear damping coefficient and lump stiffness, respectively. With mechanical pump applied, the intrinsic energy dissipation situation has been changed due to the mechanical pump-induced energy transfer processes [2], which can be described by considering the intrinsic $\zeta$ as an effective damping coefficient $\zeta_{eff}$, thus the Eq. S12 changes to

$$m\ddot{z}(t) + \zeta_{eff} \dot{z}(t) + kz(t) = F_{th} \tag{S13}$$

Using Fourier transform, the transfer function of Eq. S13 can be obtained as

$$H(\omega) = \frac{1}{-m\omega^2 + i\zeta_{eff}\omega + k} \tag{S14}$$

thus the power spectral density (PSD) of $z(t)$ is calculated by [3]

$$S_{zz}(\omega) = |H(\omega)|^2 \cdot S_{th}(\omega) = \frac{1}{(k - m\omega^2)^2 + \zeta_{eff}^2 \omega^2} S_{th}(\omega) \tag{S15}$$

where $S_{th}(\omega)$ is the PSD of thermomechanical noise force which is typically regarded as the white noise with uniform spectral density, i.e. $S_{th}(\omega) = 4k_B T \zeta$ ($k_B$ and $T$ are Boltzmann constant and temperature respectively) [2, 4]. As the damping coefficients can be calculated by

$$\zeta = \frac{m\omega_0}{Q}, \quad \zeta_{eff} = \frac{m\omega_0}{Q_{eff}} \tag{S16}$$

where $\omega_0$ is the resonance frequency given by $\omega_0 = \sqrt{k/m}$, then substituting Eq. S16 into Eq. S15, the PSD of the oscillator displacement under the excitation of thermomechanical noise force can be expressed as

$$S_{zz}(\omega) = \frac{4k_B T \omega_0 / (mQ)}{\left(\omega_0^2 - \omega^2\right)^2 + \left(\frac{\omega \omega_0}{Q_{eff}}\right)^2} \tag{S17}$$

For the experimentally measured noise PSD, usually a signal output gain *G* and a noise floor $N_{floor}$ should be considered, thereby the measured noise PSD can be represented as

$$S_{zz}^{exp}(\omega) = \frac{B \cdot \omega_0}{\left(\omega_0^2 - \omega^2\right)^2 + \left(\frac{\omega \omega_0}{Q_{eff}}\right)^2} + N_{floor}$$

$$B \equiv \frac{4k_B TG}{mQ} \tag{S18}$$

As $B$ and $N_{floor}$ are constants, fitting the measured noise PSD by Eq. S18 can extract the effective $Q$ factor.

## S4. Resonance Frequency of Mode 1 as A Function of Blue Pump Amplitude

By using Lorentz fitting to the measured noise spectra shown in Figure 4g (main text), the resonance frequency of mode 1 under different pump strength can be obtained, which is plotted as a function of pump amplitude in Figure S2. It is clear that the resonance frequency of mode 1 can almost keep constant with increasing the pump amplitude.

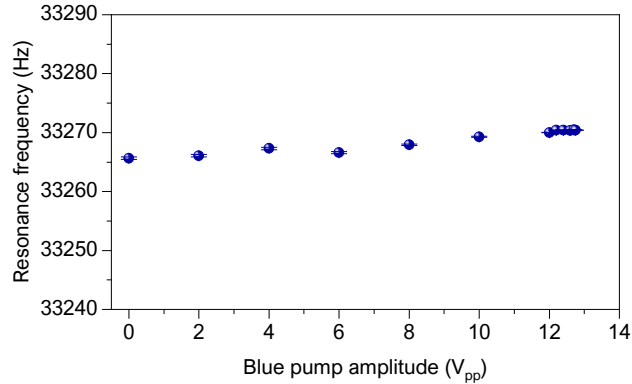

**Figure S2.** The resonance frequency of mode 1 as a function of blue pump amplitude.

## S5. The Measurement of FM-KPFM Frequency Shift under Mechanical Pump

The FM-KPFM tested here is based on the classic frequency modulation scheme [5], where a phase-locked loop (PLL) equipped with a voltage-controlled oscillator (VCO) is employed to measure the electrostatic force-induced resonance frequency shift. The mode 1 of the microcantilever is mechanically excited by the drive signal from VCO, and at the same time, the mechanical pump (here using blue pump mainly) is applied to tune the dynamic properties of mode 1. In order to study the surface potential sensitivity and SNR under mechanical pump, the FM-KPFM is operated in open-loop mode here and the basic experimental set-up is shown in Figure S3. When a conductive tip is located near the sample surface, the electrostatic force between the tip and sample is $F_{es} = C'\Delta V^2/2$, where $C'$ is the tip-sample differential capacitance, and $\Delta V$ is the total potential difference determined by the surface potential ($V_{sp}$) and DC bias ($V_{dc}$), i.e. $\Delta V = V_{dc} - V_{sp}$. Since the measurement of $V_{sp}$ in FM-KPFM is based on the $\Delta f$ signal [6], here we use the $\Delta f$ as the indicator to compare the potential sensitivity under different pump conditions. Firstly, the parabolic $\Delta f$-$V_{dc}$ curve is measured to obtain the $V_{sp}$ [7], and then this $V_{sp}$ is used as the bias offset to set the desired $\Delta V$ thus the $\Delta f$ signal with known $\Delta V$ can be attained and the potential sensitivity can be analyzed accordingly. Figure 5a of the main text is the $\Delta f$ signal measured by applying an alternating $\Delta V$ of 0 and 2 V under different mechanical pump strength.

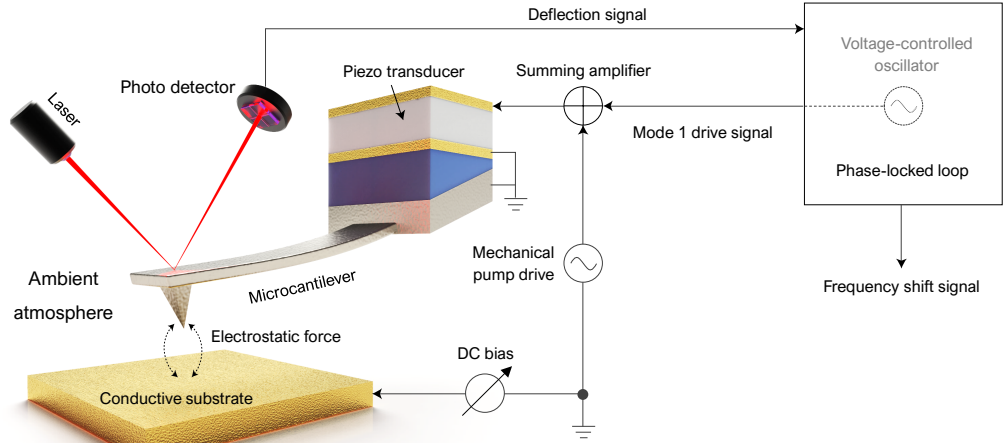

**Figure S3.** Experimental set-up for measuring FM-KPFM frequency shift under mechanical pump.

## S6. FM-KPFM Frequency Shift Signal for Calculating SNR

To calculate the SNR of $\Delta f$ signal, the noise signal $\Delta f_{noise}$ and the target $\Delta f$ signal are recorded at $\Delta V = 0$ and 2 V for a period of time respectively, and then the mean-squared values of $\Delta f_{noise}$ and $\Delta f$ can be simply extracted from the acquired data shown in Figure S4.

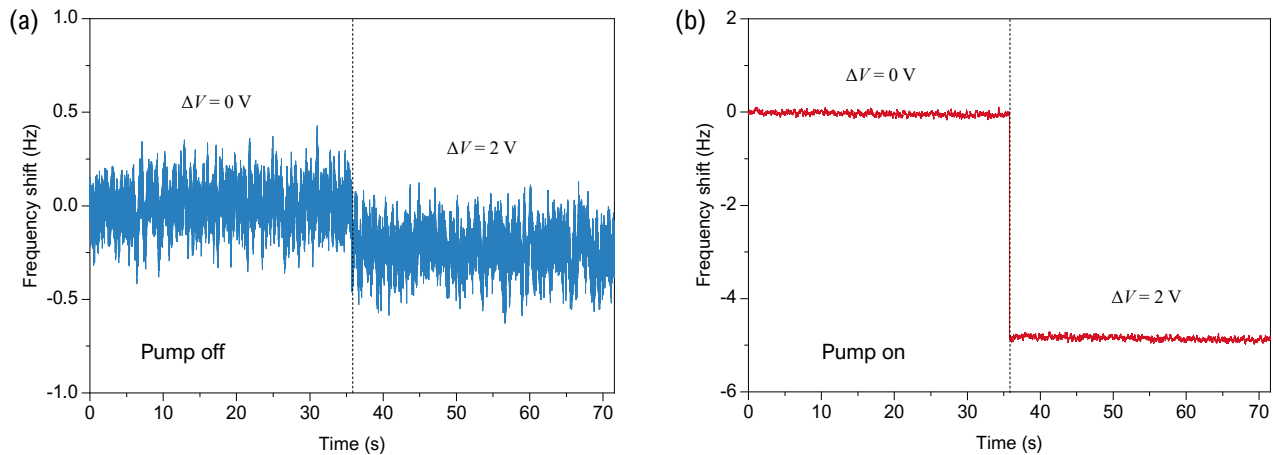

**Figure S4.** FM-KPFM frequency shift signal with and without applying mechanical pump. (a) The frequency shift signal measured at $\Delta V = 0$ and 2 V with pump off and (b) pump on.

## References

(1)  Mahboob, I.; Nishiguchi, K.; Okamoto, H.; Yamaguchi, H., Phonon-Cavity Electromechanics. *Nat. Phys.* **2012,** *8*, 387-392.